\documentclass[a4paper]{article}

\usepackage{icrc2013}
\usepackage[english]{babel}
\usepackage{amssymb}

\title{Towards the ASTRI mini-array}

\shorttitle{ASTRI mini-array performance}

\authors{
F.~Di~Pierro$^{1a}$,
C.~Bigongiari$^{1a}$,
C.~Morello$^{1a}$,
A.~Stamerra$^{1a}$,
P.~Vallania$^{1a}$,
G.~Agnetta$^{2}$,
L.A.~Antonelli$^{3}$,
D.~Bastieri$^{4}$,
G.~Bellassai$^{5}$,
M.~Belluso$^{5}$,
S.~Billotta$^{5}$,
B.~Biondo$^{2}$,
G.~Bonanno$^{5}$,
G.~Bonnoli$^{6}$,
P.~Bruno$^{5}$,
A.~Bulgarelli$^{7}$,
R.~Canestrari$^{6}$,
M.~Capalbi$^{2}$,
P.~Caraveo$^{8}$,
A.~Carosi$^{3}$,
E.~Cascone$^{9}$,
O.~Catalano$^{2}$,
M.~Cereda$^{6}$,
P.~Conconi$^{6}$,
V.~Conforti$^{7}$,
G.~Cusumano$^{2}$,
V.~De~Caprio$^{9}$,
A.~De~Luca$^{8}$,
A.~Di~Paola$^{3}$,
D.~Fantinel$^{10}$,
M.~Fiorini$^{8}$,
D.~Fugazza$^{6}$,
D.~Gardiol$^{1b}$,
M.~Ghigo$^{6}$,
F.~Gianotti$^{7}$,
S.~Giarrusso$^{2}$,
E.~Giro$^{10}$,
A.~Grillo$^{5}$,
D.~Impiombato$^{2}$,
S.~Incorvaia$^{8}$,
A.~La~Barbera$^{2}$,
N.~La~Palombara$^{8}$,
V.~La~Parola$^{2}$,
G.~La Rosa$^{2}$,
L.~Lessio$^{10}$,
G.~Leto$^{5}$,
S.~Lombardi$^{3}$,
F.~Lucarelli$^{3}$,
M.C.~Maccarone$^{2}$,
G.~Malaguti$^{7}$,
G.~Malaspina$^{6}$,
V.~Mangano$^{2}$,
D.~Marano$^{5}$,
E.~Martinetti$^{5}$,
R.~Millul$^{6}$,
T.~Mineo$^{2}$,
A.~Mist\`{o}$^{6}$,
G.~Morlino$^{11}$,
M.R.~Panzera$^{6}$,
G.~Pareschi$^{6}$,
G.~Rodeghiero$^{10}$,
P.~Romano$^{2}$,
F.~Russo$^{2}$,
B.~Sacco$^{2}$,
N.~Sartore$^{8}$,
J.~Schwarz$^{6}$,
A.~Segreto$^{2}$,
G.~Sironi$^{6}$,
G.~Sottile$^{2}$,
E.~Strazzeri$^{2}$,
L.~Stringhetti$^{8}$,
G.~Tagliaferri$^{6}$,
V.~Testa$^{3}$,
M.C.~Timpanaro$^{5}$,
G.~Toso$^{8}$,
G.~Tosti$^{12}$,
M.~Trifoglio$^{7}$,
S.~Vercellone$^{2}$,
V.~Zitelli$^{13}$
(the ASTRI Collaboration) and the CTA Consortium.
}

\afiliations{
$^{1}$ INAF - Osservatorio Astrofisico di Torino; (a) Sede di Torino - Via P. Giuria 1, 10125 Torino, Italy; (b) Sede di Pino Torinese - Strada Osservatorio 20, 10025 Pino Torinese (TO), Italy\\
$^{2}$ INAF - Istituto di Astrofisica Spaziale e Fisica Cosmica di Palermo, Via U. La Malfa 153, 90146 Palermo, Italy\\
$^{3}$ INAF - Osservatorio Astronomico di Roma, Via Frascati 33, 00040 Monte Porziocatone (RM), Italy\\
$^{4}$ Universit\`{a} di Padova, Dip. Fisica e Astronomia, Via Marzolo 8, 35131 Padova, Italy\\
$^{5}$ INAF - Osservatorio Astrofisico di Catania, Via S. Sofia 78, 95123 Catania, Italy\\
$^{6}$ INAF - Osservatorio Astronomico di Brera, Via E. Bianchi 46, 23807 Merate (LC), Italy\\
$^{7}$ INAF - Istituto di Astrofisica Spaziale e Fisica Cosmica di Bologna, Via P. Gobetti 101, 40129 Bologna, Italy\\
$^{8}$ INAF - Istituto di Astrofisica Spaziale e Fisica Cosmica di Milano, Via E. Bassini 15, 20133 Milano, Italy\\
$^{9}$ INAF - Osservatorio Astronomico di Capodimonte, Salita Moiariello 16, 80131 Napoli, Italy\\
$^{10}$ INAF - Osservatorio Astronomico di Padova, Vicolo Osservatorio 5, 35122 Padova, Italy\\
$^{11}$ INAF - Osservatorio Astrofisico di Arcetri, Largo E. Fermi 5, 50125 Firenze, Italy\\
$^{12}$ Universit\`{a} di Perugia, Dip. Fisica, Via A. Pascoli, 06123 Perugia, Italy\\
$^{13}$ INAF - Osservatorio Astronomico di Bologna, Via Ranzani 1, 40127 Bologna, Italy\\
}

\email{f.dipierro@ifsi-torino.inaf.it}

\abstract{
The Cherenkov Telescope Array (CTA) will consist of an array of three types of telescopes covering a wide energy range, from tens of GeV up to more than 100 TeV.
The high energy section ($\gtrsim$ 3 TeV) will be covered by the Small Size Telescopes (SST).
ASTRI (\textit{Astrofisica con Specchi a Tecnologia Replicante Italiana}) is a flagship project of the Italian Ministry of Research and Education led by INAF, aiming at the design and construction of a prototype of the Dual Mirror SST. In a second phase the ASTRI project foresees the installation of the first elements of the SST array at the CTA southern site, a mini-array of 5-7 telescopes.
The optimization of the layout of this mini-array embedded in the SST array of the CTA Observatory has been the object of an intense simulation effort.
In this work we present the expected mini-array performance in terms of energy threshold, angular and energy resolution and sensitivity.
}


\begin{document}
\maketitle
%
%%Begin a section.
%
\section{Introduction}

The Cherenkov Telescope Array (CTA) \cite{bib:CTA1,bib:CTA2} is a world wide project aimed at designing, building and operating an array of $\sim 100$ telescopes covering an energy range of about 4 orders of magnitude with a sensitivity in the core energy region (around 1 TeV) of about one order of magnitude better than current arrays. Due to the different features of the Cherenkov signals for energies spanning from some tens of GeV to 100 TeV and beyond, three kinds of telescopes have been proposed. At low energy the photon density is very low, so very large reflective surfaces are needed to collect enough photons to correctly reconstruct the Cherenkov images requiring a minimum number of 50-100 photoelectrons (p.e.). On the other hand, the higher incoming flux at lower energies reduces the need of large effective areas. In this energy region a small (3-4) number of Large Size Telescopes (LSTs $\sim24\,$m diameter) is envisaged. In the medium energy regime, the drop of incoming flux due to the slopes of the source spectra requires a larger collection area, built up with $\sim30$ Medium Size Telescopes of moderate aperture (MSTs, $\sim12\,$m diameter). In the highest energy range the photon density on ground is very high even for large core distances, while the source flux becomes very low. An effective area of $1-10\,$km$^2$ is thus necessary, and it is obtained placing $\sim$70 Small Size Telescopes (SSTs, $\sim4\,$m diameter) on a grid with $\sim300\,$m spacing.

The optimization of the individual telescope parameters and their layout is done using Monte Carlo (MC) simulations. 
The main figure of merit is the sensitivity, i.e. the minimum detectable flux in a given time interval, but also the angular and energy resolutions and the sensitivity across the field of view (field flatness) have to be considered for some specific physics cases. 
These performance parameters have been derived using existing and custom simulation and analysis programs. Due to the excellent background rejection achievable with the IACT stereo technique a huge amount of primary hadrons (mainly protons) has to be generated. 

For a full sky coverage the CTA project will have an array in each hemisphere. The site of the Southern array will be decided within this year (2013) and the array construction is expected to start in 2016. 

ASTRI (\textit{Astrofisica con Specchi a Tecnologia Replicante Italiana}) \cite{bib:pare1, bib:astri} is a flagship project of the Italian Ministry of Education, University and Research strictly linked to the development of CTA. Within this framework, INAF is currently developing an end-to-end prototype of the CTA small-size telescope in a dual-mirror configuration (SST-2M) to be tested under field conditions at the INAF ``M.C. Fracastoro''observing station in Serra La Nave (Mount Etna, Sicily)
\cite{bib:macc}, and scheduled to start data acquisition in 2014. Furthermore the ASTRI collaboration proposed the installation of a small array consisting of 5-7 SSTs \cite{bib:pare2} at the selected CTA Southern site, possibly supported by one or two MSTs and one LST. This small array could constitute the first CTA seed and could be used for technical purposes but also for scientific studies, mainly in the highest energy region from few tens of TeV. This energy region is widely unexplored and may yield exciting and unexpected discoveries. 

\section{Monte Carlo simulations}

The simulation and analysis of the Cherenkov images on the detector focal plane produced by the primary particles impacting on the top of the atmosphere is done with the simulation chain used to evaluate the performance of the whole CTA array and involves several steps: 
\begin{itemize}

\item generation of the Extensive Air Showers (EAS) and of the Cherenkov photon yield with CORSIKA v.6.99 \cite{bib:corsika};

\item simulation of the telescope optics and detector electronics with \textit{sim\_telarray} \cite{bib:simtel};

\item reduction and analysis of the simulated events through different reconstruction programs derived by the software used by the HESS, MAGIC and VERITAS collaborations.

\end{itemize}

The available data include a gamma point source and diffuse protons simulated within a cone of $\pm 10^{\circ}$ in order to correctly take into account the contribution of background events falling well outside the field of view. The energy ranges of the simulated events are 0.003-330 TeV and 0.004-600 TeV for gammas and protons, respectively, with differential spectral index $\gamma=-2$ (rescaled to $-2.57$ and $-2.7$ in the analysis step). The site parameters (atmospheric depth and magnetic field) have been chosen to match one of the CTA South candidate sites (El Leoncito, 2660 m a.s.l.). The showers have been scattered over a circular area with radius of 2500~m for gammas (10 times each shower) and 3000~m for protons (20 times) with observing zenith angle $\theta = 20^{\circ}$. 

Table~\ref{tab:MCsamples} summarizes the main characteristics of the samples of simulated EAS used in this work, a fraction ($\sim$10$\%$) of the full production \cite{bib:prod2} (at the moment still ongoing) which will allow more accurate results.

\begin{table}[h]
\begin{center}
\begin{tabular}{|l|c|c|}
\hline  & Gamma &  Proton \\ \hline
Core position re-sampling   &  10  &  20 \\ \hline
Events (after re-sampling) &  2.6e8  &  4.5e9  \\ \hline
Energy [TeV]  &  0.003 - 330  &  0.004 - 600 \\ \hline
Radius  [m]   &  2500  &  3000 \\ \hline
\end{tabular}
\caption{Monte Carlo samples.}
\label{tab:MCsamples}
\end{center}
\end{table}

Among the different telescopes proposed for SST, in this analysis the dual mirror telescopes \cite{bib:pare2} have been used. Table~\ref{tab:sstconf} shows the main features for this configuration.

\begin{table}[h]
\begin{center}
\begin{tabular}{|l|c|}
\hline 
Primary mirror (M1) diameter [m]  &  4.3   \\ \hline
Secondary mirror (M2) diameter [m]  & 1.8    \\ \hline
Distance Focal plane - M2 [m]  & 0.52    \\ \hline
Distance M1 - M2 [m]  & 3.1    \\ \hline
\end{tabular}
\caption{Characteristics of the simulated SST-2M.}
\label{tab:sstconf}
\end{center}
\end{table}

\begin{figure}[htb]
  \centering
  \includegraphics[width=0.5\textwidth]{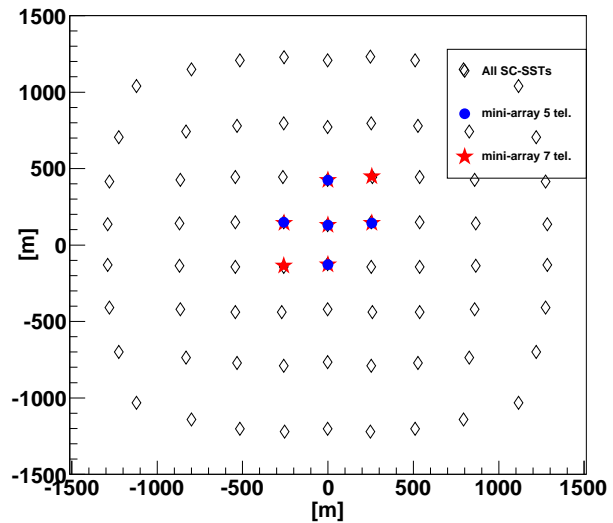}
  \caption{Selected sub-arrays within the whole SST-2M array.}
  \label{fig:layout}
 \end{figure}

The study of the mini-array performance is done selecting subsets of 5 and 7 SSTs from the whole layout as shown in Figure~\ref{fig:layout}.
 
Data are then reconstructed and analyzed with different programs \cite{bib:prod2} and their results are cross-checked. All the results from the different analyses are compatible within uncertainties of the order of 20-30\%. In the following results based on the MARS \cite{bib:mars} and read\_hess \cite{bib:simtel} programs are shown.
 
\subsection{Effective area and threshold energy}

\begin{figure}[htb]
  \centering
  \includegraphics[width=0.5\textwidth]{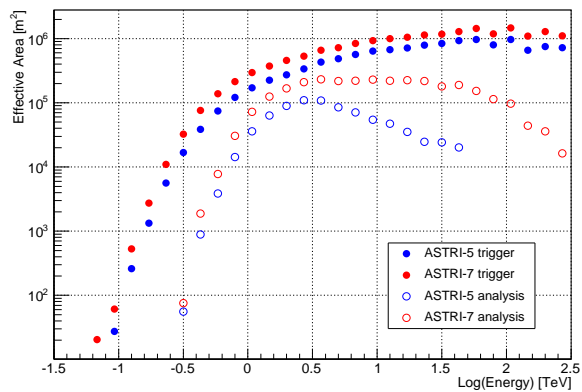}
  \caption{Effective area for gammas at trigger level and after shape cuts.}
  \label{fig:effareag}
 \end{figure}

\begin{figure}[htb]
  \centering
  \includegraphics[width=0.5\textwidth]{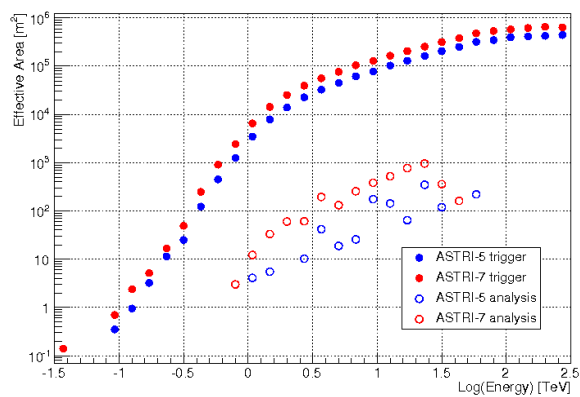}
  \caption{Effective area for protons at trigger level and after shape cuts.}
  \label{fig:effareap}
 \end{figure}

Figures~\ref{fig:effareag} and \ref{fig:effareap} show the effective area after trigger (i.e.: at least 2 telescopes) and after analysis cuts, not including the arrival direction one, for $\gamma$ and protons. As expected, due to the smaller shower size for proton-generated EAS with respect to gammas of the same energy, the trigger effective area for protons is smaller than for gammas. The 7 telescopes layout provides an effective area $\sim50\%$\ larger than the 5 telescopes's one. 
 
From the trigger effective area and an assumed Crab-like source spectrum \cite{bib:crab} we derive a trigger energy threshold around 500 GeV (corresponding to the peak). The energy threshold for analysis obtained from the effective area after analysis cuts is around 1 TeV. Since the smallest distance between telescopes is 260 m for both the 5 and the 7 telescopes sub-arrays, the two configurations have the same trigger threshold.

\subsection{The angular and energy resolutions}

\begin{figure}[htb]
  \centering
  \includegraphics[width=0.5\textwidth]{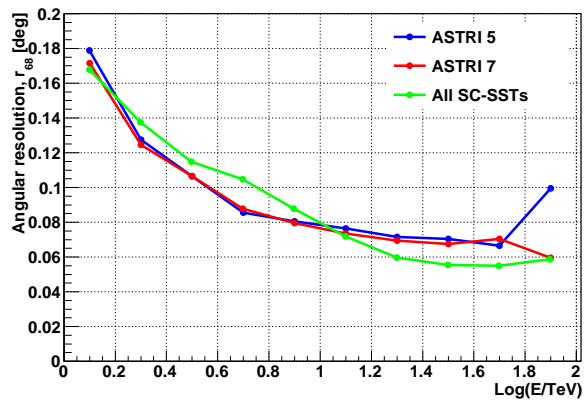}
  \caption{Angular resolution for 5 and 7-fold mini-array configurations and the full SC-SST array.}
  \label{fig:angres}
 \end{figure}

The aim of the CTA project is to boost not only the sensitivity but also the angular and energy resolutions. Figure~\ref{fig:angres} shows the expected angular resolution at 68\% containment for the mini-arrays and for the whole SST array of CTA, with at least two images used in reconstruction. The requirement on the number of images at low energies (E $<$ 10 TeV) results in the selection of a higher quality subsample for the mini-arrays with respect to the full SST. On the other hand at high energies the angular resolution benefits of the higher multiplicity.    

\begin{figure}[htb]
  \centering
  \includegraphics[width=0.5\textwidth]{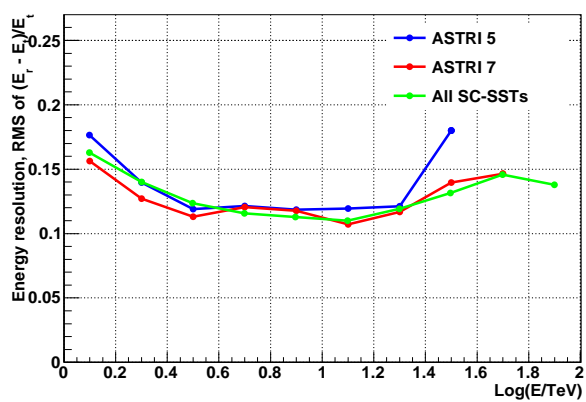}
  \caption{Energy resolution for 5 and 7-fold mini-array configurations and the full SC-SST array.}
  \label{fig:eneres}
 \end{figure}

The energy resolution, shown in Figure~\ref{fig:eneres}, which can be as good as 10\% at a few TeV, approaches the full SST result and is in any case limited by the intrinsic shower-to-shower fluctuations.

Despite the much lower effective area of the mini-array with respect to the CTA array, the expected angular and energy resolutions will be comparable at least for certain data selections and energies. Checking on-field the mini-array performance will be a check of the whole SST array and a validation of the entire simulation procedure.

\subsection{The gamma/hadron discrimination and sensitivity}

\begin{figure*}[htbp]
  \centering
  \includegraphics[width=0.9\textwidth]{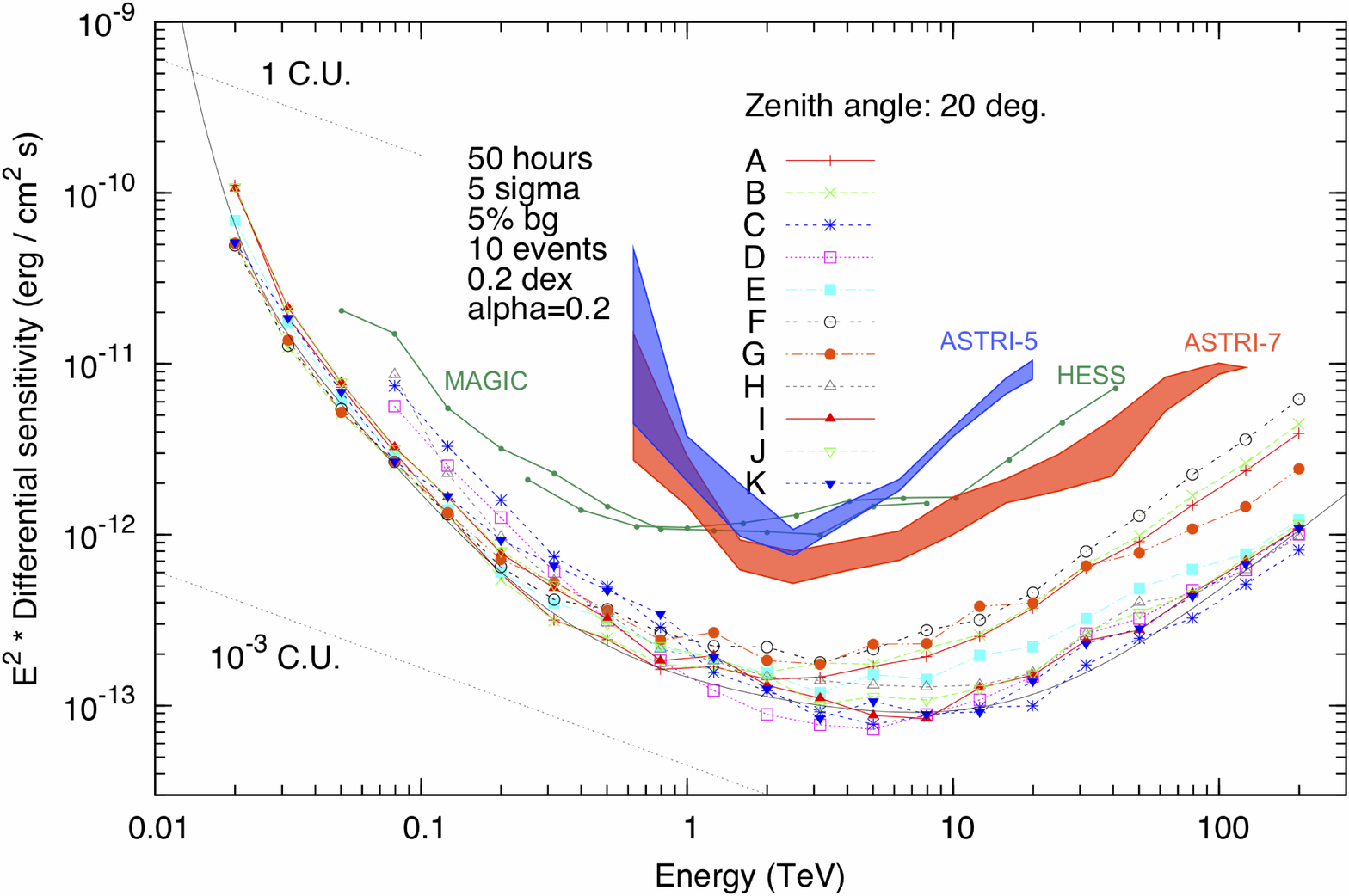}
  \caption{Comparison of differential sensitivities for different explored layouts for CTA \cite{bib:CTAProd1}, HESS, MAGIC and this work mini-arrays (5 and 7 SST-2M).}
  \label{fig:sensitivity}
 \end{figure*}

The gamma/hadron discrimination has been optimized for point sources selecting cuts on the image shape and on the shower direction, based on the well known fact that gamma-generated images are more compact with respect to proton ones and that the major axis is directed towards the source direction.
From the effective areas after the $\gamma$/hadron cuts the expected sensitivity as a function of the primary energy can be calculated for a standard observing time of 50\,h and a five standard deviations statistical significance. For the significance calculation, equation 17 of Li\&Ma\cite{bib:LiMa} has been used assuming a background region 5 times larger than the signal region ($\alpha$ = 0.2). The other requirements are: 5 bins/decade in energy; a minimum of 10 excess events above the background in each energy bin; the signal excess at least 5\% of the remaining background after cuts in order to take into account possible systematics in background subtraction. Figure~\ref{fig:sensitivity} shows the expected differential sensitivity compared to the whole CTA (several explored layouts) and HESS and MAGIC. While at low energy HESS and MAGIC perform clearly better, starting from $E=10\,$TeV the mini-array expected sensitivity, for the 7 telescopes layout, is better up to a factor 2 at few tens of TeV and is still sensitive at 100 TeV, where the effective area of present generation IACTs drops dramatically. In this energy regime the mini-array will operate as the most sensitive IACT array.

\section{Discussion and conclusions}

The quick installation of a mini-array of SSTs in the selected CTA site will be of paramount benefit for the whole project. 
The comparison between expected performance with experimental ones will be a validation of the MC chain for the whole SST. The ASTRI mini-array will allow to test several innovative adopted technical solutions such as the Dual Mirror approach and the SiPM-based camera. Another important test will be the wide field technique, i.e. the observation of the higher energy events very far from the core (up to ~500\,m) and outside the Cherenkov light pool. Due to the limited field of view of actual detectors this technique has never been tested in the field, and this can easily be done with at least 2 telescopes. For the layout optimization a specific grid with telescope separation between 200 and 350\,m must be used and dedicated simulations are ongoing. The addition of two MSTs and possibly one LST would complete the low energy range allowing the inter-calibration checks between different sub-arrays. Finally, from the scientific point of view, the ASTRI mini-array will be the detector with the best performance for energies greater than 10\,TeV. This regime is largely unexplored and could provide interesting results on known sources but also unexpected ones~\cite{bib:verc}.

\vspace*{0.5cm}
\footnotesize{{\bf Acknowledgment:}{
This work was partially supported by the ASTRI Flagship Project financed by the Italian Ministry of Education, University, and Research (MIUR) and led by the Italian National Institute of Astrophysics (INAF). We also acknowledge partial support by the MIUR Bando PRIN 2009. We gratefully acknowledge support from the agencies and organizations   
  listed in this page: http://www.cta-observatory.org/?q=node/22}.}

\begin{thebibliography}{}

\bibitem{bib:CTA1} M. Actis et al., Exp. Astr. 32 (2011) 193-316; arXiv 1008.3703.
\bibitem{bib:CTA2} B.S. Acharya et al., Astroparticle Physics 43 (2013) 3.
\bibitem{bib:pare1} G. Pareschi et al., [The ASTRI Collaboration], (in preparation).
\bibitem{bib:astri} http://www.brera.inaf.it/astri
\bibitem{bib:macc} M.C. Maccarone, et al., id. 0110, these proceedings.
\bibitem{bib:pare2} G. Pareschi et al., id. 0466, these proceedings. 
\bibitem{bib:corsika} D. Heck et al., Forschungszentrum Karlsruhe Report No. FZKA 6019 (1998).
\bibitem{bib:simtel} K. Bernl\"ohr, Astroparticle Physics 30 (2008) 149.
\bibitem{bib:prod2} K. Bernl\"ohr et al., id. 1053, these proceedings. 
\bibitem{bib:mars} A. Moralejo et al., Proc. 31st ICRC (2009).
\bibitem{bib:crab} Aharonian et al., The Astrophysical Journal 614 (2004) 897.
\bibitem{bib:LiMa} T. Li \& Y. Ma, The Astrophysical Journal 272 (1983) 317.
\bibitem{bib:CTAProd1} K. Bernl\"ohr  et al., Astroparticle Physics 43 (2013) 171.
\bibitem{bib:verc} S. Vercellone et al., id. 0109, these proceedings.


%\bibitem{bib:schoenberg} M. Sch\"onberg and S. Chandrasekhar, The Astrophysical Journal 96 (1942) 161-172 doi:10.1086/144444.
%
\end{thebibliography}
\end{document}